\documentclass[aps,prb,twocolumn,showpacs,preprintnumbers,amsmath,amssymb,tightenlines,epsfig]{revtex4}

\usepackage{amsmath}
\usepackage{graphicx}
\usepackage{dcolumn}
\usepackage{bm}

\begin{document}

\title{Real Time Relativity: exploration learning of special relativity}
\author{C. M. Savage,  A. Searle, and L. McCalman}
\affiliation{Department of Physics, 
Australian National University,
ACT 0200, Australia}

\email{craig.savage@anu.edu.au}

\begin{abstract}
``Real Time Relativity'' is a computer program that lets students fly at
relativistic speeds though a simulated world populated with planets, clocks,
and buildings.  The counterintuitive and spectacular optical effects of
relativity are prominent, while systematic exploration of the simulation
allows the user to discover relativistic effects such as length contraction and the relativity of simultaneity.  We report on the physics and technology underpinning the simulation, and our experience
using it for teaching special relativity to first year university students.
\end{abstract}

\maketitle

\section{Introduction}
\label{Introduction}

\begin{figure}
\includegraphics[width=\columnwidth]{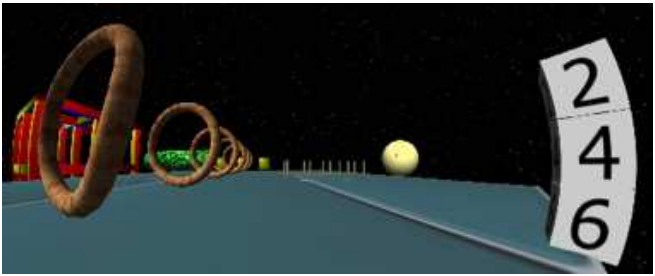}
\caption{Screenshot from Real Time Relativity. The speed of the camera relative to the objects is $v = 0.9682c$. The Doppler and headlight effects have been turned off.}
\label{fig:RTR1}
\end{figure}

``Real Time Relativity''  \cite{RTR} is a first person point of view game-like computer simulation of a special relativistic world, which allows the user to move in three dimensions amongst familiar objects, Fig.~\ref{fig:RTR1}. In a first year university physics course it has proved complementary to other relativity instruction.
%

Since there is little opportunity for students to directly experience relativity, it is often perceived as abstract, and they may find it hard to form an integrated relativistic world-view. They find relativity interesting and exciting, but may be left bemused by the chasm between the theory and their everyday experience. \cite{Scherr 2001,Scherr 2002,Scherr thesis} Real Time Relativity can help bridge this chasm by making visual observations the basis from which the theory is deduced.

In his original relativity paper Einstein discarded the personal observer, who collects information from what he sees, in favor of more abstract inertial observers who use distributed arrays of rulers and conventionally synchronized clocks. \cite{Einstein} However Komar \cite{Komar} and others \cite{Peres,Blatter} showed that special relativity may be formulated in terms of postulates about a personal observer's visual observations. This  approach to relativity underpins our use of the Real Time Relativity simulation.

Studies have shown that students may fail to learn fundamental concepts, such as the relativity of simultaneity, even after advanced instruction. \cite{Scherr 2001,Scherr 2002,Scherr thesis} This is because special relativity contradicts some deeply held ideas about space and time. To overcome their non-relativistic preconceptions students must first recognize them, and then confront them.  \cite{Scherr 2002} The Real Time Relativity simulation can aid this process.

In the next section we discuss some relevant physics education research. Section \ref{Relativistic Optics} outlines the relativistic optics required to understand the simulation. Section \ref{Technology} briefly overviews the computer technology that is making interactive simulations of realistic physics, such as Real Time Relativity, increasingly practical. Section \ref{The Real Time Relativity Simulation} describes students' experience of the simulation.  Section \ref{The Relativity of Simultaneity} shows how it provides fresh perspectives on physics such as the relativity of simultaneity. Section \ref{Laboratory evaluation} reports our evaluation of its use in a first year physics laboratory.

\section{Educational background}
\label{Educational background}

There is substantial evidence for the value of active learning. \cite{Hake,PER-1}  Effective learning is stimulated by students participating in the construction and application of physics based world-views.  \cite{McDermott 2001} A common factor in active learning is a cycle of developing, testing, and correcting understanding in a collaborative environment. Peer instruction is one way for this to occur in the classroom, \cite{Mazur} while in the laboratory inquiry based approaches are known to be effective. \cite{McDermott 2001}

Computer simulations may promote active learning in physics, especially where real laboratories are difficult to provide. However, research has shown that a testing and development cycle is required to ensure good learning outcomes.  \cite{Wieman Nature Physics,Adams} The effectiveness of simulations is reduced by poor interfaces, \cite{Adams} and by students' lack of the skills required to learn from them. \cite{Yeo} They also lose effectiveness if the exploration is not conducted according to the scientific method. \cite{Steinberg} However, when such issues are addressed the results can be spectacular. \cite{Wieman Nature Physics}

Computer simulations are most effective when directed towards clear goals, with an understanding of their strengths and limitations. \cite{Wieman Nature Physics,Steinberg} They can provide an additional active learning mode, and address broad goals such as ``thinking like a physicist''. \cite{Van Heuvelen} However, learning to use software increases cognitive load, lessening capacity to learn other new material. \cite{Yeo} The value for physics teaching of first person simulations, such as Real Time Relativity, is largely unexplored, as existing research has concerned simulations of models, such as might be used by an expert physicist, rather than immersive first person simulations. \cite{Wieman Nature Physics,Steinberg,Yeo,Adams}

Real Time Relativity differs from other physics simulations in providing a realistic, explorable environment.  In the context of a first year university physics class,  we are asking the question: Can aspects of special relativity be learnt by exploration of the Real Time Relativity virtual world? Many students are comfortable interactively discovering the rules of virtual worlds; perhaps they can use this experience for discovering the rules of physics?

Successful learning from simulations is more likely if students are suitably prepared and guided.  \cite{Yeo,McDermott 2001,Adams} Preparation should develop a basic understanding of  the physics which determines what is seen in the simulation. In our case this includes the finite speed of light, the Doppler effect, and relativistic optical aberration.  This preparation might use conventional interactive multimedia. \cite{TEE} Preparation should also include how the scientific method is used to develop understanding of novel phenomena.

Scherr, Shaffer, and Vokos have found that students' understanding of time in special relativity is poor. \cite{Scherr 2001,Scherr 2002} They conclude that ``... many students who study special relativity at the undergraduate to graduate levels fail to develop a functional understanding'' . \cite{Scherr 2002} They  identify the reason for this as students misunderstanding fundamental ideas such as: the ``time of an event, simultaneity, and reference frame''.  \cite{Scherr 2002} They have developed instructional materials to address these problems.  Mermin \cite{Mermin} has also noted that traditional relativistic pedagogy may make incorrect assumptions about students' prior knowledge. Real Time Relativity can address these problems, as fundamental ideas, such as the time of an event, have intuitive operational meanings.

\section{Relativistic Optics}
\label{Relativistic Optics}

Some of the basic physics of relativistic optics, namely the Doppler effect and aberration, was discussed by Einstein in his first relativity paper. \cite{Einstein} However it was not until about 1960 that the pioneering work of Penrose, \cite{Penrose} Terrell,  \cite{Terrell} and Weisskopf \cite{Weisskopf} showed that relativity gives a rich and unexpected visual environment. 

In this section we summarize relativistic optics using 4-vectors, because that is how it is implemented in the Real Time Relativity program (see Section \ref{Technology}). Rindler \cite{Rindler} provides a more complete introduction, both with and without using 4-vectors.

A plane light wave is described by its 4-frequency $F$, which has components \cite{Rindler}
\begin{equation}
F  = f [1,\vec{n}] ,
\label{4-frequency-light}
\end{equation}
where $f$ is the frequency and  $\vec{n} = (n_x, n_y, n_z)$ is the unit vector in the propagation direction. 
From the 4-frequency components in a particular frame, the components in any other frame may be found using a Lorentz transformation. The transformation between the usual standard configuration \cite{standard configuration} frames $S$ and $S'$ are sufficient for our purposes. We will use ``world'' ($w$) and ``camera'' ($c$) to refer to the frames $S$ and $S'$,
\begin{eqnarray}
&f_c =  \gamma f_w ( 1 - n_{w,x} v/c ) ,
\nonumber \\
&f_c n_{c,x}  = \gamma f_w ( n_{w,x} -  v/c ) ,
\nonumber \\
&f_c n_{c,y} = f_w n_{w,y}, \; \;  f_c n_{c,z} = f n_{w,z} ,
\nonumber \\
& \gamma = (1 -v^2 / c^2 )^{-1/2} ,
\label{4-frequency LT}
\end{eqnarray}
where $v$ is the component of the relative velocity of the frames along the positive $x_w$ axis of the world frame. The first equation expresses the Doppler effect, and the remaining equations express the dependence of the propagation direction on the relative frame velocity: an effect known as ``relativistic aberration''. In the Real Time Relativity simulation let the world frame be that in which the objects are at rest, and the camera frame be the user's instantaneous rest frame, as we represent the user by a  camera. We require the frequencies and propagation directions in the camera frame, $f_c$ and $\vec{n}_c$.

Since $\vec{n}_c$  is a unit vector, its $x$-component, $n_{c,x}$ is the cosine of the angle  $\theta_c$ between the light ray and the $x_c$ axis: if the ray is coming towards the observer $n_{c,x}$ changes sign.
Dividing the second and third of Eqs.~(\ref{4-frequency LT}) by the first and using $n_{c,x} = - \cos \theta_c$ and $n_{w,x} = - \cos \theta_w$,  we get
\begin{eqnarray}
&\cos \theta_c = \frac{\cos \theta_w +v/c}{1+(v/c)\cos \theta_w}  ,
\nonumber \\
&\sin \theta_c = \frac{\sin \theta_w }{\gamma (1+(v/c)\cos \theta_w ) }  .
\label{Aberration angles}
\end{eqnarray}

Relativistic aberration is analogous to non-relativistic forms of aberration that students may have experienced. For example, the dependence of the angle of falling rain on an observer's velocity, or the difference between the visual position of a high flying aircraft and that indicated by its sound. This understanding may be made quantitative using the relativistic velocity addition formulae. \cite{Rindler}

\begin{figure}
\includegraphics[width=\columnwidth]{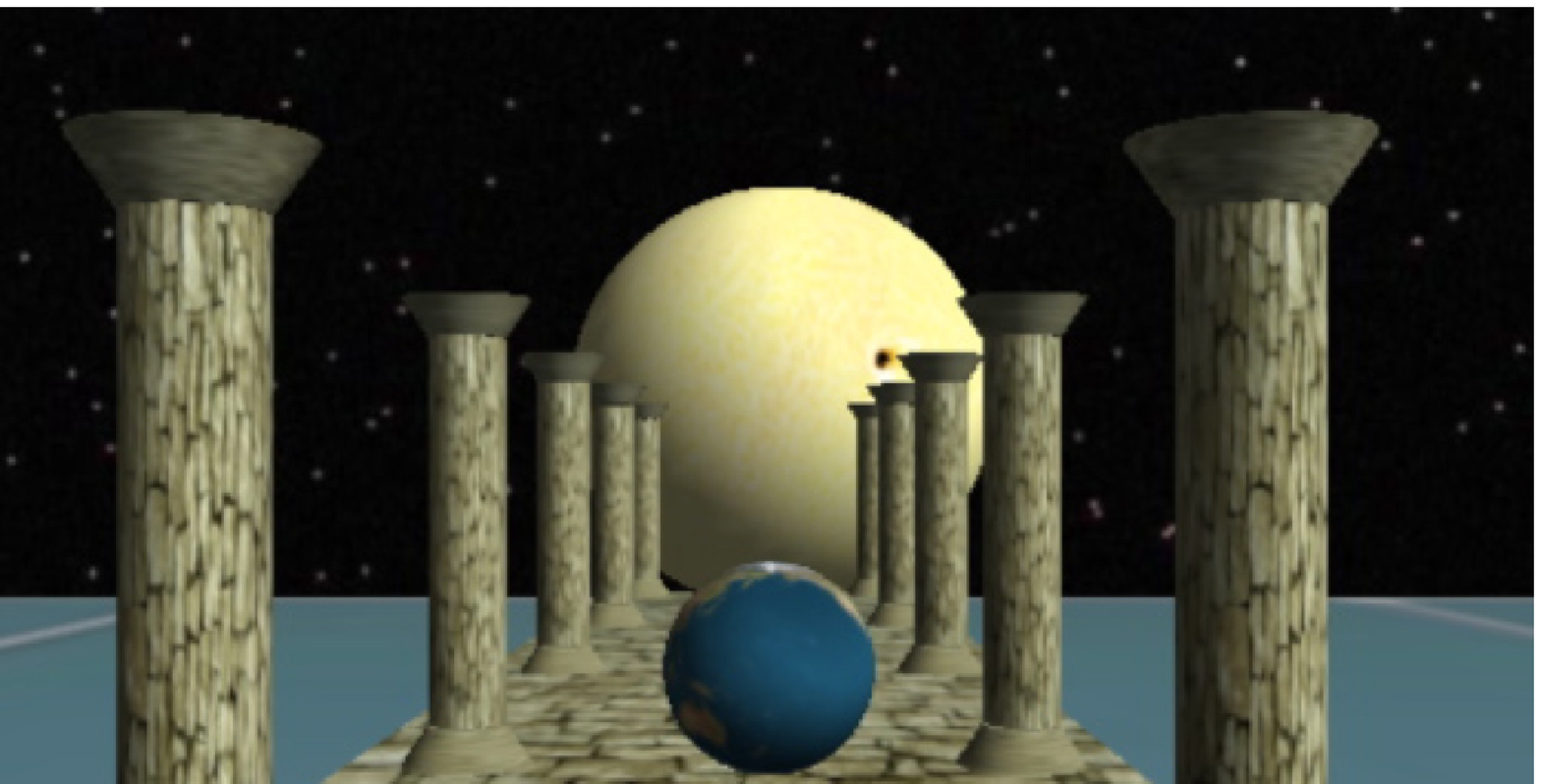}
\includegraphics[width=\columnwidth]{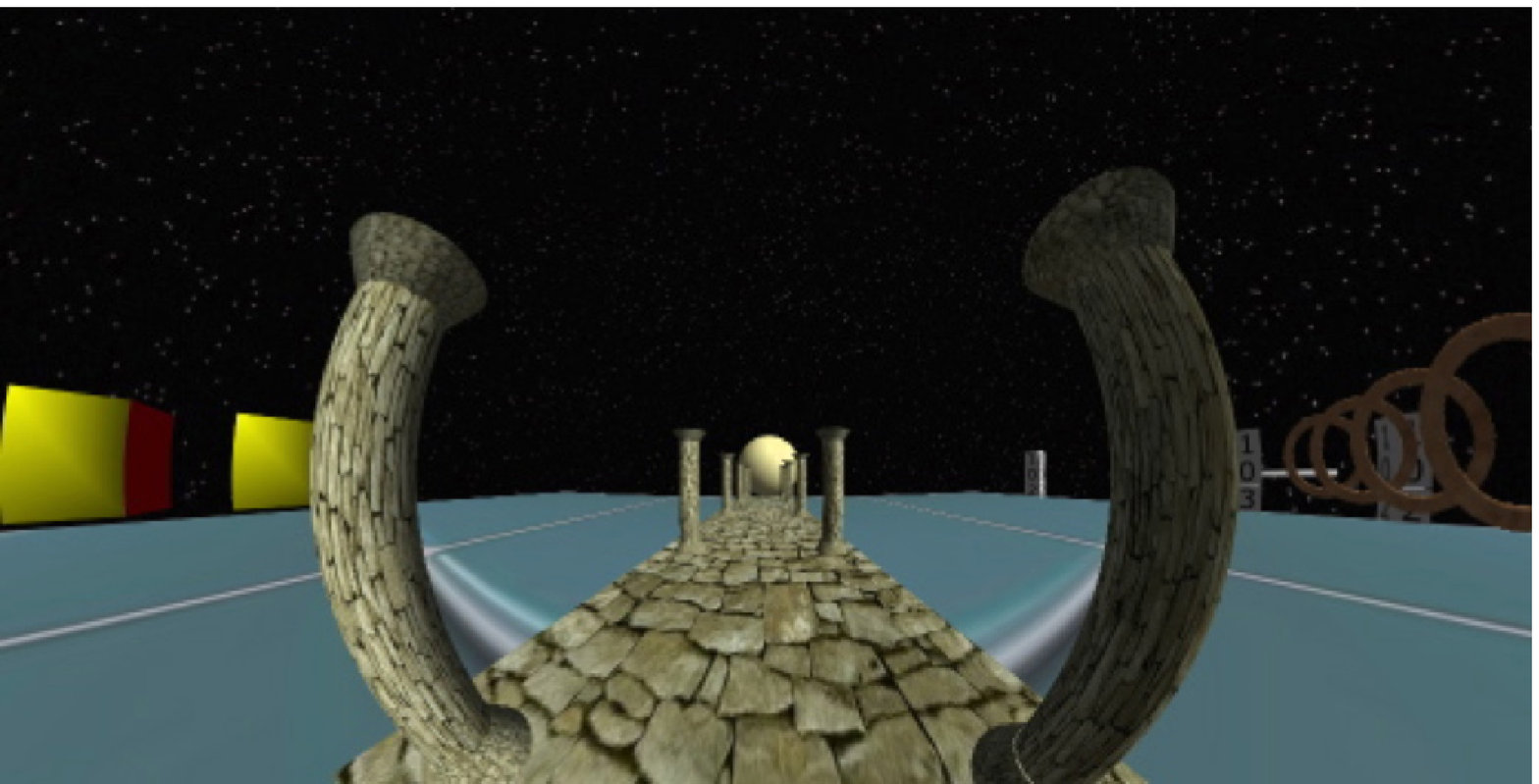}
\caption{Screenshots from Real Time Relativity. Top frame: at rest in the world frame. Bottom frame: $v = 0.9682 c$, corresponding to $\gamma = 4$. The Doppler and headlight effects have been turned off. In the world frame the camera is in front of the position in the top frame. }
\label{fig: RTR start}
\end{figure}

Penrose \cite{Penrose} showed that relativistic aberration implies that straight-lines  are seen as either lines or circular arcs in other frames.  He also showed that a sphere, which always has a circular outline (unlike a circle which may have an elliptical outline), will continue to have a circular outline after aberration, and hence continue to look like a sphere. These effects are immediately apparent in the Real Time Relativity simulation, Fig.~\ref{fig: RTR start}.
%

The non-relativistic Doppler effect may also be familiar to some students. This, together with the analogy to non-relativistic aberration,  emphasize the closer relation of relativistic optics to direct experience than the usual space-time approach to special relativity.

A convenient form of the Doppler effect follows from the first two of Eqs.~(\ref{4-frequency LT}) after eliminating $n_{w,x}$,
\begin{equation}
f_c = \frac{\sqrt{1-v^2/c^2}}{1 - (v/c) \cos \theta_c} f_w = D f_w ,
\label{Doppler fdash angles}
\end{equation}
where this equation defines the ``Doppler factor'' $D$. For $ v/c \ll 1$ the denominator is the familiar non-relativistic wave compression or expansion. For waves incoming at   $\theta_c = \pi/2$ radians to the relative motion, the denominator is one and the observed frequency is less than the world frequency, at which they were emitted. This means that the time between wave crests, the period, is longer; which is exactly the effect of time dilation, if the wave crests are regarded as a clock.

The effect of aberration on small angles may be found by taking differentials of the inverse Lorentz transformation of the 4-frequency. \cite{ILT} This yields
\begin{equation}
d \theta_C = D^{-1} d \theta_W .
\label{angle ratio}
\end{equation}
Hence small angles transform by the inverse Doppler factor. In particular, for objects directly ahead, so that $\theta_C = 0$, and for $v/c \ll 1$, the inverse Doppler factor is approximately $D \approx 1-v/c$, and objects angular sizes are shrunk. While for objects behind, $D \approx 1+v/c$ and objects are expanded.

Perhaps the most subtle of the relativistic optics effects is the headlight effect. Indeed, a complete discussion was not given until 1979, by McKinley. \cite{McKinley starbow, McKinley power} It refers to the increased intensity of light coming from objects we are moving towards. The intensity decreases for objects we are moving away from. Three things combine to produce  these intensity changes: the change in angular size of the emitting region, the Doppler change in energy of the photons, and the change in photon flux due to the combined effects of time dilation and the observer's motion, which is an additional manifestation of the Doppler effect. In terms of the Doppler factor in Eq.~(\ref{Doppler fdash angles}) these contribute factors to the change in intensity of $D^2$, $D$, and $D$ respectively, for a combined intensity change factor of $D^4$. However for common detectors, such as the eye or a CCD camera, it is the photon number flux $P$ that is detected, and this changes by a factor of $D^3$, since the energy change per photon is irrelevant,
\begin{equation}
P_B  = D^3 P_A  .
\label{photon flux}
\end{equation}

\section{Technology}
\label{Technology}

Computers can generate images incorporating special, \cite{Chang} and general, \cite{Muller} relativistic optics. By the early 1990s it was possible to interactively render simple objects, such as cubes. \cite{Gekelman} The highest quality images were generated by the ray-tracing method, which is capable of producing photo-realistic images. \cite{Hsiung} However ray-tracing is currently too slow for interactive simulations, although individual images can be strung together to make movies. \cite{TEE}

The development of the programmable graphics processor  \cite{Owens} has made it possible to render complex relativistic scenes in real time. The first such systems appear to have been developed by D. Weiskopf \cite{Weiskopf} and M. Bochers  \cite{Bochers} within the physics education group at the University of T\"ubingen. \cite{Tubingen web site} This group has focussed on using relativity visualization for science communication. \cite{Weiskopf IEEE} Our Real Time Relativity simulation is similar to these, is freely available, and is being developed as an Open Source project under the Lesser General Public License.  \cite{RTR} 

The screen image displayed by Real Time Relativity is created using the computer graphics technique known as environment mapping, which renders the three-dimensional virtual world onto a two-dimensional cube map. A cube map may be visualised as the $4 \pi$ sterradian view-field mapped onto the interior surface of a cube centered on a camera, representing the user's field of view. In fact, the cube map is a data structure in which the image pixels are addressed by line of sight direction, rather than by spatial position. The relativistically correct scene is produced by transforming the cube map.

Each camera image pixel is formed by light incident from a particular direction; that is, with a specific propagation direction $\vec{n_c}$ in the camera frame. The relativistic physics problem is to find the corresponding propagation vector $\vec{n_w}$ in the world frame in which the cube map is constructed. This vector then addresses the pixel on the cube map that is mapped to the camera pixel. The resulting camera image is displayed on the screen.

A plane light wave is represented by the relativistic 4-frequency, Eq.~(\ref{4-frequency-light}). The propagation direction in the world frame is found by the inverse Lorentz transformation \cite{ILT} of this 4-vector from the camera frame into the world frame. This is implemented as a four-dimensional matrix multiplication of the 4-frequency. The transformation matrix is calculated before each frame is rendered, using the current camera velocity, and is then applied to a 4-frequency constructed for each camera pixel. This has a spatial component equal to the pixel's imaging direction and the time component set to one. The spatial part of the transformed 4-frequency addresses the cube map pixel that is then rendered to the screen. 

The Graphics Processing Units (GPUs) on computer video cards are designed to do 4-vector matrix algebra efficiently and in parallel. This makes it possible for simulations such as Real Time Relativity, and that due to Bochers, \cite{Bochers,Weiskopf IEEE} to perform the Lorentz transformations in real time. The 4-vectors that graphics processing units normally work with specify the $x, y$ and $z$ coordinates of a vertex and a fourth $w$ component that facilitates certain non-linear transformations (such as translation and perspective projection), or specify the red, green, blue colour and alpha (transparency) of a (texture) pixel.  Since the processing of different vertices or pixels is usually independent, the operations can be performed in parallel.

The Doppler shift factor $D$  is given by the ratio of the time components of the 4-frequencies in the camera and world frames, Eq.~(\ref{Doppler fdash angles}). However, to determine the effect of the Doppler shift on a general colour requires the entire intensity spectrum. But in current graphics systems the spectrum is specified at just three frequencies; red, green, and blue. Hence interpolation is used to generate the spectrum. This simple approach, together with the lack of any infrared or ultraviolet spectra, prevents a true representation of Doppler shifted colors, and is a significant limitation of the current version of Real Time Relativity. In particular,  stars do not maintain a blackbody spectrum. \cite{McKinley starbow,Greber,Kraus}

The headlight effect, Eq.~(\ref{photon flux}), is implemented by multiplying each pixel color vector by the third power of the Doppler shift factor $D$. There are significant limitations on how the resulting large intensity range is rendered to the screen by current hardware.

The graphics processing unit does the Lorentz transformations as well as its usual graphics work. First, a non-relativistic three-dimensional scene is rendered to a cube map, then relativistic transformations are applied to it.  To generate a frame, the 4-frequency associated with each camera pixel is inverse Lorentz transformed to find the corresponding world frame cube map pixel. This is then Doppler and intensity shifted, also by the graphics processing unit. An 800 by 600 window has 480,000 pixels, so displaying 50 frames per second requires 24 million pixel transformations per second, which is well within the capabilities of inexpensive graphics processing units. Consequently, it is the conventional graphics processing generating the cube map that limits the overall performance, not the relativistic calculations.

Real Time Relativity is programmed using Microsoft's DirectX 9 interface, so that it is independent of the details of any particular graphics processing unit. Consequently, it is only available on Windows computer systems. DirectX 9 includes the High Level Shader Language in which the pixel shader controlling the graphics processing unit is written.

Graphics processing units have been increasing in processing power more rapidly than central processing units. \cite{Owens} This is driven by the demand for parallel computing from the gaming community.  For example, the Xbox 360 graphics processing unit has forty-eight 32-bit processors running at 500 MHz, each capable of a floating point 4-vector operation per cycle, giving nearly 100 GigaFlops, compared to perhaps a few GigaFlops for a central processing unit. \cite{Xbox} The main limitation is that graphics processing units do data-parallel computing, in which the same operation is repeated on each element of a data array. Nevertheless, computational scientists are developing algorithms that harness their processing power for tasks such as solving partial differential equations.  \cite{Owens} The Folding@Home distributed computing project has a client available which runs their molecular dynamics calculation on graphics processing units, increasing computational power by about twenty times per computer. \cite{Folding at home} These developments may have an impact on the kinds of physics teaching simulations that are possible in the future.

\section{The Real Time Relativity Simulation}
\label{The Real Time Relativity Simulation}

In this section we introduce the Real Time Relativity simulation as experienced by students in the first year course for physics majors at The Australian National University. \cite{PHYS1201} It was used in a three hour laboratory structured to encourage exploration, while requiring that certain measurements be made and compared to theory. Students were provided with a manual giving background information and asking both qualitative and quantitative questions.  Many students completed the laboratory before they attended the relativistic optics lecture. Students worked in groups of two or three, and discussion was encouraged. Preparation included answering simple pre-lab questions which were assessed at the beginning of the laboratory.

An initial problem of orientation within relativistic optics simulations arises because the speed of light is very large in everyday terms. This means that either the objects in the simulation must be very large, roughly light-seconds, or the speed of light must be artificially slow, as in Gamow's Mr.~Tompkins story. \cite{Gamow} In the interest of realism, we have taken the former view, which allows us to include realistic astronomical objects such as the Earth, which is 0.042 light-seconds in diameter. The top frame of Fig.~\ref{fig: RTR start} shows a screen from Real Time Relativity. The Earth is visible, as is the Sun behind it. \cite{sun} These objects set the scale of the simulated world. Other objects, such as the columns, have been chosen for their familiar shapes, although they would be absurdly large if they existed in the real world. Familiar objects aid in the recognition of the distortions caused by relativistic aberration.

Students start by accelerating from rest down the row of columns shown in Fig.~\ref{fig: RTR start}. At first  it seems that they are moving backwards. \cite{movies} This is completely counterintuitive and prompts them to question what they see: the exploration has begun! The effect is due to relativistic aberration. An important way that motion is sensed is by the change in angular size as our distance to the object changes. Normally, as we approach an object its angular size increases, roughly proportionate with the distance. In contrast, the {\it decrease} in angular size due to relativistic aberration, Eq.~(\ref{angle ratio}), is approximately proportional to $1-(v/c)\cos \theta_c$, for $v/c \ll 1$, and occurs before the distance has time to change. Therefore the initial view is of objects shrinking, and this is interpreted as them moving away, and hence as backwards movement of the viewer. As they continue to accelerate, nearby objects eventually pass by, and the perception of forward motion is restored. 
%
\begin{figure}
\includegraphics[width=\columnwidth]{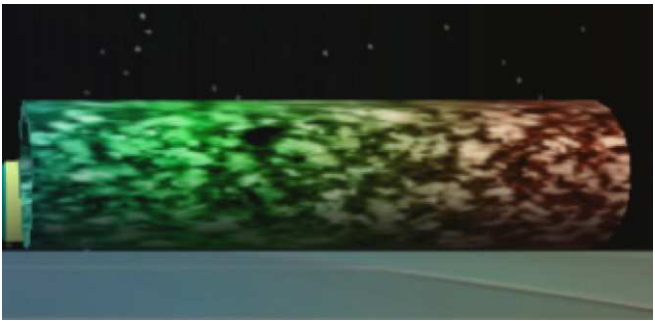}
\caption{Screenshot from Real Time Relativity showing how the Doppler effect depends on the view angle. The camera is looking perpendicular to the direction of motion, which is from right to left at $v = 0.5 c$, at a stippled blue cylinder. The Doppler factor is approximately one, $D \approx 1$, in the direction of the left edge of the image. The rest of the cylinder is  red shifted through green and red. The headlight effect has been turned off.}
\label{fig: RTR rainbow}
\end{figure}

Colors change due the Doppler effect Eq.~(\ref{Doppler fdash angles}), but the headlight effect quickly saturates the scene with bright light, dominating all other effects, due to its dependence on the third power of the Doppler factor, Eq.~(\ref{photon flux}). Consequently, it is useful to be able to turn it off. Although this goes against the principle of making the simulation as realistic as possible, it is difficult to see some other effects if it is left on.

The Doppler effect  depends on the viewing angle, Eq.~(\ref{Doppler fdash angles}). There is a particular angle to the direction of motion $\theta_0$ for which there is no effect, since the Doppler factor $D=1$ when
\begin{equation}
\cos \theta_0 = (c/v) ( 1- \gamma^{-1} ) .
\label{no doppler angle}
\end{equation}
For $v=0.5c$ this angle is $ \theta_0 = 1.3$ radians ($74^\circ$). If a student looks at at a pure colored object at this angle they see a rainbow effect, as for directions towards that of motion, $\theta < \theta_0$, the color is blue shifted, while for directions away from the direction of motion, $\theta > \theta_0$, the color is red shifted. Fig.~\ref{fig: RTR rainbow} shows the red shifting of a blue cylinder through green and red as the viewing angle increases. 

If students already understand the non-relativistic Doppler effect, they may be guided to discover the relativistic version. In particular, it is possible to deduce time dilation from the observation that there is reddening when viewing perpendicular to the direction of motion, Eq.~(\ref{Doppler fdash angles}).

The bottom frame of Fig.~\ref{fig: RTR start} shows the scene with the camera travelling down the row of columns with a speed corresponding to $\gamma = 4$. The Doppler and headlight effects have been turned off. The circular curvature of the nearest columns is due to relativistic aberration, as discussed in section \ref{Relativistic Optics}. The curvature of the more distant columns is barely noticeable. However, they are shrunk by approximately the inverse Doppler factor $D^{-1}$, according to Eq.~(\ref{angle ratio}). The camera field of view covers a wide field in the world frame: the hoops and cubes on the edges of the image are behind the camera in the world frame. 

\begin{figure}
\includegraphics[width=\columnwidth]{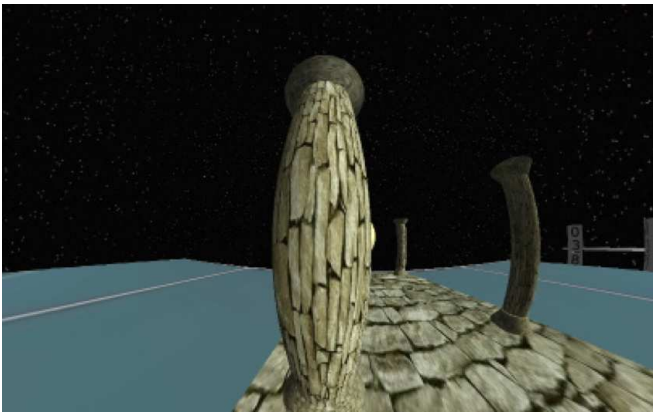}
\caption{Screenshot from Real Time Relativity showing relativistic aberration.The camera is moving towards the center of the column with $v=0.9682c$, so that $\gamma = 4$. The Doppler and headlight effects are off. }
\label{fig: RTR column head on}
\end{figure}

Aberration may be understood as a consequence of the finite speed of light. The key idea is that the light that reaches the camera at a particular instant was reflected by objects at different times.  The light from closer objects was reflected later than that from far away objects. This is irrelevant when the camera is at rest relative to the objects, but when it is moving, the position of the objects in the camera frame depends on time. For a large object this means that the parts nearer the camera reflected the received light later than the further parts. If the camera is moving towards the object, at a significant fraction of the speed of light, the near parts reflect when they are significantly closer and hence look bigger than the far parts which reflected when they were further away and hence looked smaller. If we are moving directly towards the middle of an object the net result is that the middle looks fatter than the ends, see Fig.~\ref{fig: RTR column head on}. If the object is off to one side it  is curved into a circular arc.
%

Currently, Real Time Relativity is limited to all objects being at rest in the world frame. This means that relativistic dynamics is not within its capabilities.

\section{The Relativity of Simultaneity}
\label{The Relativity of Simultaneity}

\begin{figure}
\includegraphics[width=\columnwidth]{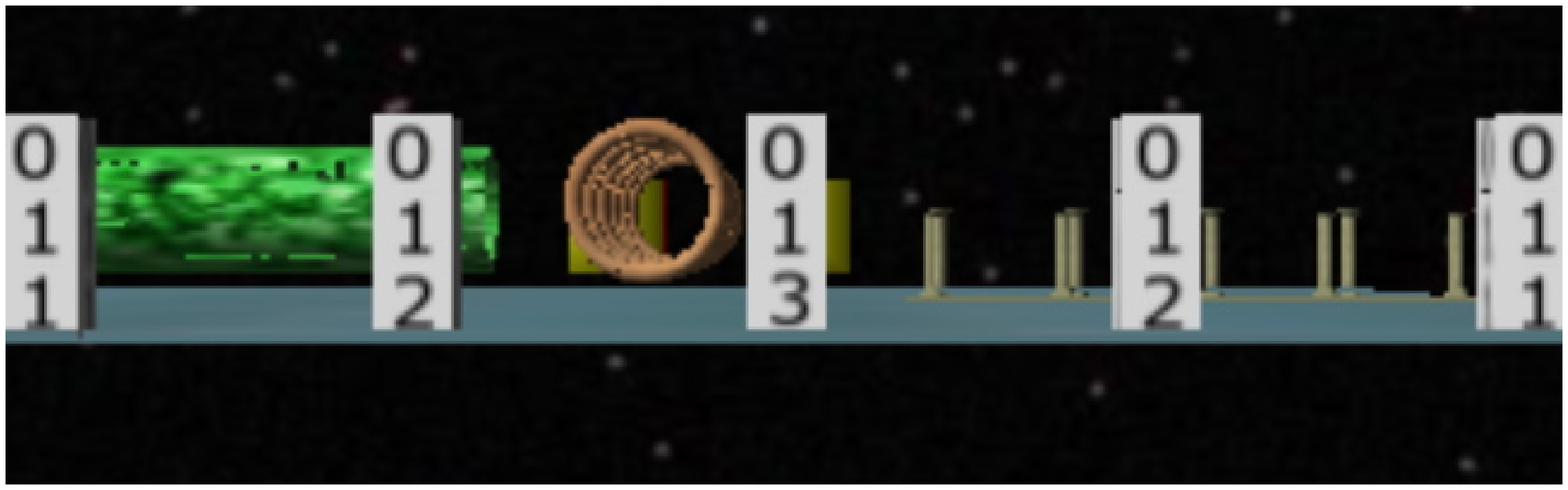}
\includegraphics[width=\columnwidth]{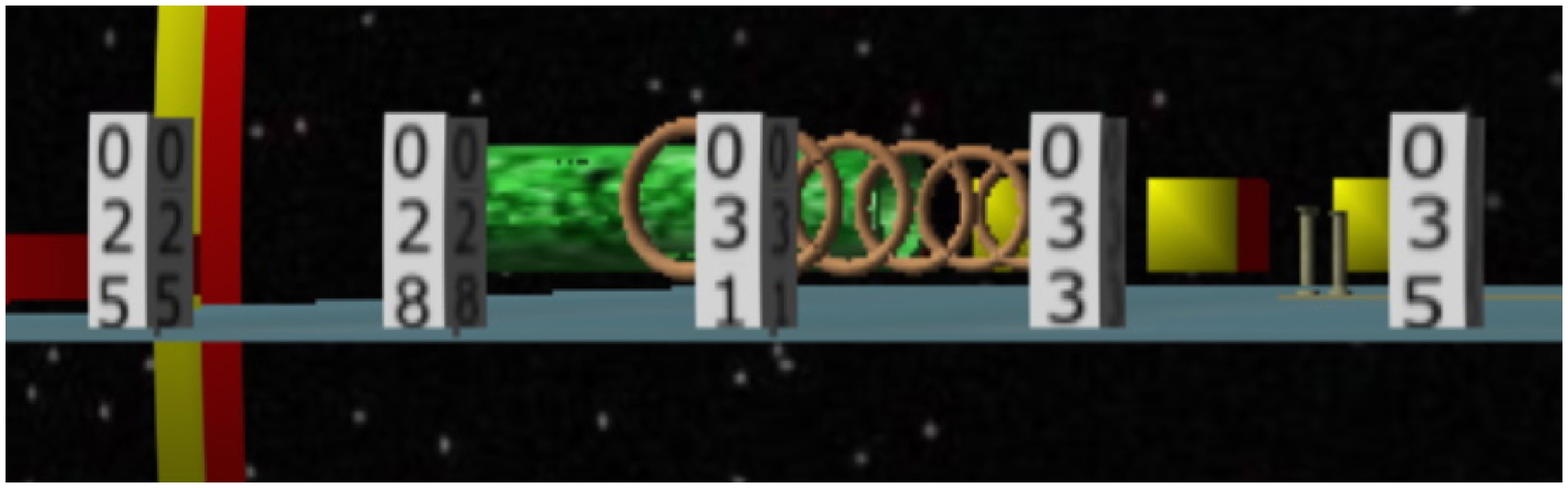}
\includegraphics[width=\columnwidth]{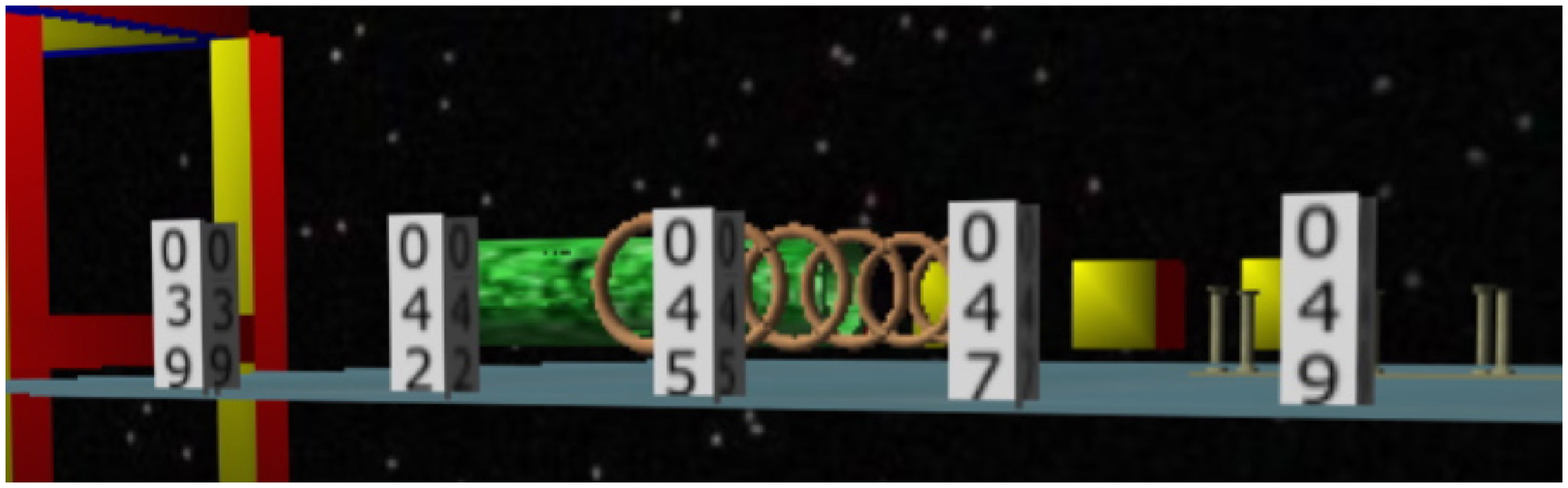}
\caption{Screenshots from Real Time Relativity explaining the relativity of simultaneity. Top frame. The effect of light propagation delay on observed clocks. The camera is at rest relative to the clocks, which are lined up perpendicular to the line of sight to the central clock. The clocks are 5 light-seconds apart and read seconds. Middle frame. The camera is moving from left to right parallel to the clocks with $v=0.5c$. The perpendicular distance to the clocks is the same as in the top frame (about 31 light-seconds). The  major contributor to the different clock readings is the relativity of simultaneity. However, light delay causes clocks to the left to differ more from the central clock than those to the right.  Bottom frame. The camera has been brought to rest immediately after taking the middle frame, although some time then elapsed before the image was taken. In the clocks' rest frame the different clock readings are entirely a consequence of the light propagation delay. The field of view is the same in each frame.}
\label{fig: RTR clocks}
\end{figure}

The relativity of simultaneity has been identified as a particularly difficult concept for students to learn from passive instruction. \cite{Scherr 2001,Scherr 2002,Scherr thesis} In order that students might actively discover the relativity of simultaneity for themselves, the Real Time Relativity simulation includes  clocks in the world frame. Even when the camera is at rest in the world frame, clocks at different distances from the camera are seen to read different times due to the light propagation delay, see the top frame of Fig.~\ref{fig: RTR clocks}. Students generally have no difficulty recognizing and utilizing this fact. \cite{Scherr 2001,Scherr 2002,Scherr thesis} Note that clocks the same distance from the camera read the same time.
%

The middle frame shows the same view of the clocks, but with the camera moving with $v=0.5c$ parallel to the clocks from left to right. The camera is looking perpendicular to its direction of motion. Note that the eye gets confusing cues from this image, as the clocks are rotated as if we were looking at them from in front, but we are not. This effect is a result of relativistic aberration known as ``Terrell rotation''.  \cite{Terrell} Length contraction by the factor $\gamma^{-1} = 0.87$ is also apparent.

The relativity of simultaneity is apparent from the readings on the clocks in the middle frame of Fig.~\ref{fig: RTR clocks}. The right-most clock is ahead of the left-most by 10 seconds. 
This cannot be explained by light delay {\it in the camera frame}, as the observed time difference is too  large, and the times increase from left to right. 
However it is explained by light delay in the clocks' frame. Students can see this by immediately stopping the camera relative to the clocks. Due to relativistic aberration they must then look back to see the clocks: this view is shown in the bottom frame of Fig.~\ref{fig: RTR clocks}. In the clocks' rest frame the camera is not opposite the clocks, but is to their right. From this perspective it is clear why the clocks read as they do: the left-most clock is furthest  and reads earliest, while the right-most is closest and reads latest. The time difference between them is exactly that seen by the moving camera.

Let us restate the argument in terms of  two photographers: Alice is moving relative to the clocks, and Bob is stationary relative to the clocks. Both Alice and Bob take photographs of the clocks at an event  ``CLICK'', chosen so that Alice, in her own frame, is approximately equidistant from the locations of the clocks when they emitted the photographed light. \cite{equidistant} Both Alice and Bob are sampling the set of photons originating from the clocks and present at CLICK. These photons carry the same information; in particular, the times read by the clocks when they were emitted. The different times of the different clocks is understood by Bob as a result of the light propagation delay over the different distances to the clocks. However, the clocks were at approximately the same distance from Alice when they emitted the light, so she requires another explanation. This is a new physical effect: the relativity of simultaneity. The relativity postulate ensures that what is true for these clocks is true for any clocks, and hence for time itself. A complete discussion is given in Appendix \ref{Appendix}.

\section{Laboratory evaluation}
\label{Laboratory evaluation}

Real Time Relativity was incorporated in a first year laboratory session at The Australian National University. \cite{PHYS1201}  The course included nine lectures and three tutorials on special relativity. One lecture was devoted to relativistic optics. 

The content of the laboratory has been indicated in section \ref{The Real Time Relativity Simulation}. Its effectiveness was assessed in three ways. First, students completed questionnaires before and after the laboratory. Second, one of the authors was present as an observer in each laboratory, recording how students interacted with the simulation. A laboratory demonstrator was also present. Third, students recorded their work in laboratory log books which were assessed.

The pre-laboratory surveys indicated that students usually had prior knowledge of relativity and were eager to learn more. However they tended to perceive it as an abstract subject. The post-laboratory surveys indicated that students felt they had learnt about relativity from the simulation, and that it had stimulated their interest. Some students reported that the ``concrete'' or ``visual'' nature of the simulation was helpful: \cite{quotes}
\begin{quote}
``Real Time Relativity is very useful - many people are visual learners.''
\end{quote}
However, students often reported that the laboratory manual was too prescriptive and did not allow them to adequately pursue their own investigations. This criticism focussed on the quantitative exercises: 
\begin{quote}
``Why are we forcing equations from the simulation?''
\end{quote}
There were also many complaints about the difficulty of using the program, and the inadequate time available to develop proficiency with navigation through the virtual world:
\begin{quote}
``The controls were really, really hard to use.''
\end{quote}

The laboratory observer enabled a testing and refinement cycle. We identified problems, and corrected them, before the next student group took the laboratory. In particular, students often tried to push simulations to the limits to see what happened, behavior noted by the University of Colorado Physics Education Technology group. \cite{Wieman Nature Physics,Adams} If a simulation does not respond sensibly, students can lose confidence in its reliability. Observers were able to monitor what engaged students, and what frustrated them. The most engaging aspect was the exploration of a novel and open ended world. Amongst the more frustrating things were the simulation's controls not behaving in ways students considered natural.

The log books completed during the laboratory did not capture the excitement that was observed in working laboratory groups. However, successful quantitative measurements were generally made: for example, of the Doppler effect and of length contraction as a function of speed. 

Our experience confirmed the importance of developing educational software through a testing and refinement cycle. \cite{Wieman Nature Physics,Adams} Students used the simulation in ways we had not anticipated, and had different ideas to the authors about what constituted a natural user interface. The flaws in the simulation had a bigger negative impact on the students than expected. Students sometimes attributed their lack of understanding of the physics of the simulation to a ``bug'', even when there was none, rather than to their need to develop better understanding.  \cite{Adams}

Our experience suggests that the Real Time Relativity simulation can stimulate discovery learning, and provide complementary learning opportunities to those provided by lectures and problem solving tutorials. However, realizing its full value will require further cycles of testing and development. Next time we use it, we shall require students to ``play'' with the simulation as part of the pre-laboratory preparation, so that they have some familiarity with the controls and with the peculiarities of navigation in a relativistic world. We shall also provide more opportunity for open ended exploration, as this appears to be its strength.

\section{Conclusion}
\label{Conclusion}

Real Time Relativity is an immersive physics simulation of a kind that is becoming increasingly accessible due to the improving cost effectiveness of computer technology. It gives students the opportunity to discover and confront their misconceptions about relativity, and to construct resolutions.

Our experience with Real Time Relativity suggests that it provides new perspectives on special relativity. This may be particularly valuable to students who prefer the concrete over the abstract. Important physics, such as the relativity of simultaneity, can be introduced with minimal mathematics. This may broaden the group of students who can learn relativity. However, the educational value of first person simulations, like Real Time Relativity, is an interesting area for further physics education research.

\appendix
\section{The relativity of simultaneity}
\label{Appendix}

In this appendix we expand on the explanation of the relativity of simultaneity in terms of light delays that was introduced in section \ref{The Relativity of Simultaneity}. It uses the aberration and length contraction formulae. In the context of discovery learning with the Real Time Relativity simulation each of these formulae may, in principle, be deduced from observations. Along the way we also deduce time dilation.

\begin{figure}

\includegraphics[width=\columnwidth]{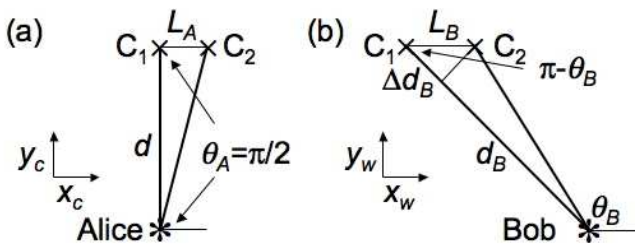}
\caption{Schematic diagrams for the relativity of simultaneity. Both panels refer to the time of event CLICK, indicated by *s, when the photographs are taken. The lines from clocks $C_1$ and $C_2$ to Alice and Bob are the paths taken by the light forming the photographs in their respective frames. (a) Alice's frame.  (b) Bob's frame in which the clocks are at rest.}
\label{fig: simultaneity schematic}
\end{figure}

We will refer to Fig.~\ref{fig: simultaneity schematic}, which shows schematic diagrams of the scenario shown in Figs.~\ref{fig: RTR clocks}. At event CLICK both Alice and Bob take photographs of the clocks. We choose CLICK to be the co-incident origins of Alice's and Bob's rest frames, which we assume to be in standard configuration with relative velocity $v$. Therefore CLICK occurs at times $t_A = t_B = 0$. In the notation of section \ref{Relativistic Optics} Alice's frame is the camera frame, and Bob's frame is the world frame.
%

Fig.~\ref{fig: simultaneity schematic}(a) shows the light paths taken from the clocks $C_1$ and $C_2$ to Alice, for whom they are moving from right to left with speed $v$. She looks perpendicular to the direction of relative motion to see them, at $\theta_A = \pi/2$, and infers that was their direction when they emitted the light she images. Let the perpendicular distance to clock $C_1$ be $d$, and the distance between the clocks, in Alice's frame, be $L_A$. Due to the light propagation delay, the time on the photograph of clock $C_1$ will be that it read at time $t_A = -d/c$. The path length difference between the paths from clocks $C_2$ and $C_1$ is
\begin{equation}
\Delta d_A =  \sqrt{d^2 + L_A^2} -d  \approx  L_A^2 /(2d) ,
\label{M path difference}
\end{equation}
where we have assumed $L_A \ll d_A$ and Taylor expanded the square root to first order. The corresponding light propagation time difference can be made arbitrarily small by making $L_A$ a sufficiently small fraction of $d \;$ \cite{equidistant}.

Fig.~\ref{fig: simultaneity schematic}(b) shows the light paths taken from the clocks to Bob, who is at rest relative to them. He looks back at the angle $\theta_B$ to photograph them. Let the distance to clock $C_1$ be $d_B$. Since lengths perpendicular to the relative motion are invariant this is given in terms of $d$ by
\begin{equation}
d_B \sin ( \pi -\theta_B) = d_B \sin \theta_B = d 
\Rightarrow
d_B = \gamma d ,
\label{O path}
\end{equation}
where we used the aberration formulae Eqs.~(\ref{Aberration angles}), with $\theta_A = \pi/2$, to find $\sin \theta_B = \gamma^{-1}$. Due to the light propagation time from $C_1$ to Bob, the time on $C_1$'s  photograph will be that it read at time $t_B = -d_B / c = -\gamma d / c$. This differs from the time deduced by Alice by the time dilation factor $\gamma$. Thus we obtain time dilation from aberration.

However the focus here is on the relativity of simultaneity. The path length difference $\Delta d_B$ between the paths from clocks $C_1$ and $C_2$ may be approximated by a method familiar from diffraction theory. We drop a perpendicular to $C_2$ from the line between clock $C_1$ and Bob. The distance along this line from the perpendicular to $C_1$ is the approximate path length difference. Using the corresponding right-angle triangle with hypotenuse $L_B$ and angle $\pi -\theta_B$ we have
\begin{equation}
\Delta d_B  = L_B \cos ( \pi - \theta_B) 
= -L_B \cos \theta_B = L_B (v/c) ,
\label{O path difference}
\end{equation}
%
where we again used the aberration formulae, Eqs.~(\ref{Aberration angles}), with $\theta_A = \pi/2$, to find $\cos \theta_B = -v/c$. The corresponding light propagation time difference, $\Delta t_B = L_B v/c^2$, is the time difference between the clocks in Bob's photograph. However, it is also the time difference between the clocks in Alice's photograph, since both images are made from the same group of photons; those present at event CLICK.

We can express this time difference in terms of Alice's quantities by using the length contraction formula $L_B =  \gamma L_A$,
\begin{equation}
\Delta t_B = (\gamma L_A ) (v/c^2) = \gamma (L_A v/c^2 ),
\label{M time difference}
\end{equation}
which is precisely the term responsible for the relativity of simultaneity in the inverse Lorentz transformation,
\begin{equation}
\Delta t_B =  \gamma ( \Delta t_A + \Delta x_A v/c^2 ) .
\label{ILT}
\end{equation}
Thus we have shown how the relativity of simultaneity can be understood in terms of light propagation delays, and be deduced from direct observations of clocks.

\begin{acknowledgments}
The authors acknowledge the assistance of Dr.~K.~Wilson with the implementation and evaluation of the Real Time Relativity laboratory.
\end{acknowledgments}



\begin{thebibliography}{99}
\bibitem{standard configuration} Standard configuration of frames $S$ and $S'$ has the $S'$ origin traveling with speed $v$ along the positive $x$ axis of $S$. The frames' origins are coincident at $t=t'=0$, and the corresponding coordinate axes are parallel.
\bibitem{ILT} The inverse Lorentz transformations of the 4-frequency are obtained from Eqs.~(\ref{4-frequency LT}) by swapping the $w$ and $c$ labels and changing the sign of $v$.
\bibitem{PHYS1201} PHYS1201. Lectures and tutorials were 50 minutes each. There were 74 students in the class.
\bibitem{Gamow} G. Gamow, {\it Mr Tompkins in Paperback} (Cambridge U.P., Cambridge, England, 1965). Note that the figures do not show what would actually be seen, as this was not known at the time the story was written.
\bibitem{sun} Although the Sun and Earth are the correct size (Sun's diameter is 4.7 light seconds) the distance between them is not to scale.
\bibitem{movies} The best way to experience this is to run the simulation. However, a movie of the sequence discussed in the text is available from the Real Time Relativity web site. \cite{RTR}
\bibitem{equidistant} Choosing the perpendicular distance to the clocks sufficiently large, compared to the distance between the clocks, makes this approximation as good as required. It is only within this approximation that pure length contraction and relativity of simultaneity are observed.
\bibitem{quotes} Quotations are taken from the post-laboratory evaluations completed by students.
\end{thebibliography}
\end{document}